\documentclass{svjour3}     
\smartqed  
\usepackage{graphicx}
\usepackage{amssymb, latexsym, textcomp,  textcomp}
\journalname{JLTP}
\begin{document}

\title{Effect of helium surface fluctuations on the Rydberg transition of trapped electrons}

\titlerunning{Effect of helium surface fluctuations on the Rydberg transition of trapped electrons}        

\author {Mikhail Belianchikov$^1$$^\dagger$, Natalia Morais$^1$, and Denis Konstantinov$^1$}


\institute{1. Quantum Dynamics Unit, Okinawa Institute of Science and Technology (OIST) Graduate University, Tancha 1919-1, Okinawa 904-0412, Japan.\\ 
}

\institute{1. Okinawa Institute of Science and Technology, Tancha 1919-1, Okinawa 904-0412, Japan \\
 \email{m.belianchikov@oist.jp} }


\maketitle
\begin{abstract}
 Electrons trapped on the surface of liquid helium is an extremely clean system which holds promise for a scalable qubit platform. However, the superfluid surface is not free from fluctuations which might cause the decay and dephasing of the electron's quantized states. Understanding and mitigating these fluctuations is essential for the advancement of electrons-on-helium (eHe) qubit technology. Some work has been recently done to investigate surface oscillations due to the mechanical vibration of the cryostat using a superconducting coplanar waveguide (CPW) resonator. In the present work, we focus on a sub-hertz frequency range and observe a strong effect of surface oscillations on the temporal dynamics of the Rydberg transition of electrons confined in a microchannel trapping device. We suggest possible origin of such oscillations and find a reasonable agreement between the corresponding estimation of the oscillation frequency and the observed result.           

\end{abstract}

\section{Introduction}
\label{intro}  

Development of quantum technologies shows an increasing importance of material selection for constructing a suitable quantum platform~\cite{lyonJofP2006,erikaNatComm2014,zhouNat2022,zhouNatPhys2024}. Among a wide variety of materials utilized in artificial quantum systems, electrons on nonpolar dielectric substrates stand out due to their exceptional cleanness~\cite{Andrei,eQLS2024}. In particular, an electron-on-helium (eHe) qubit stands out due to its potential for high spin coherence and scalability~\cite{lyonPRA2006,BradPRL2011,erikaPRA2023}. It was proposed that manipulation of the spin state of an electron can be facilitated by coupling to its motional states~\cite{shusterPRL2010,erikaAPL2024}. However, the surface of superfluid helium is not free from disturbance that can cause decoherence of the qubit. A recent experiment on the coupling of the in-plane motional state of a single trapped electron to a superconducting coplanar waveguide (CPW) resonator revealed an unusually large transition linewidth of about 80~MHz, which is almost two orders of magnitude larger than what is expected from the decay rate due to the dominant two-ripplon emission process~\cite{koolNatComm2019}. The main contribution to the observed electron linewidth was attributed to dephasing from classical helium fluctuations induced by the mechanical vibrations of the pulse tube (PT) cryocooler used to operate a ``dry'' dilution refrigerator. Such fluctuations were detected as a slowly varying jitter of the CPW resonator frequency, with the most spectral density lying below 10~Hz~\cite{koolNatComm2019}. Using another method with a higher cut-off frequency for the noise spectra measurements, Beysengulov {\it et al.} concluded that PT vibrations are responsible only for about 20\% of the rms amplitude of helium surface displacement, while the main contribution comes from the frequency range 30-60~Hz due to building vibrations. Both studies, however, did not extend below 1~Hz in the investigated noise spectra. 

In this work, we focus on sub-hertz frequency range which is probed by observing the temporal dynamics of the transition between out-of-plane quantized (Rydberg) states of eHe confined in a microchannel device~\cite{Rees2011,Rees2012,Ikeg2010,Ikeg2012,Rees2016,linJLTP2019}. In such a setup, the electric field exerted on eHe perpendicular to the helium surface due to the image charges induced in the surrounding electrodes and applied bias voltages is sensitive to the depth of liquid helium in the channel, and this affects the Rydberg transition frequency via the Stark effect. The observed oscillation of the helium depth with the period of about 7~s can be attributed to the inertia of superfluid film flow inside the experimental cell. Since helium surface vibrations could be detrimental to the coherence of the Rydberg state, proper countermeasures must be employed to reduce or completely eliminate such effects.        

\section{Experimental method}
\label{sec:method}

\begin{figure}[htt]
\centering
	\includegraphics[width=1.0\textwidth]{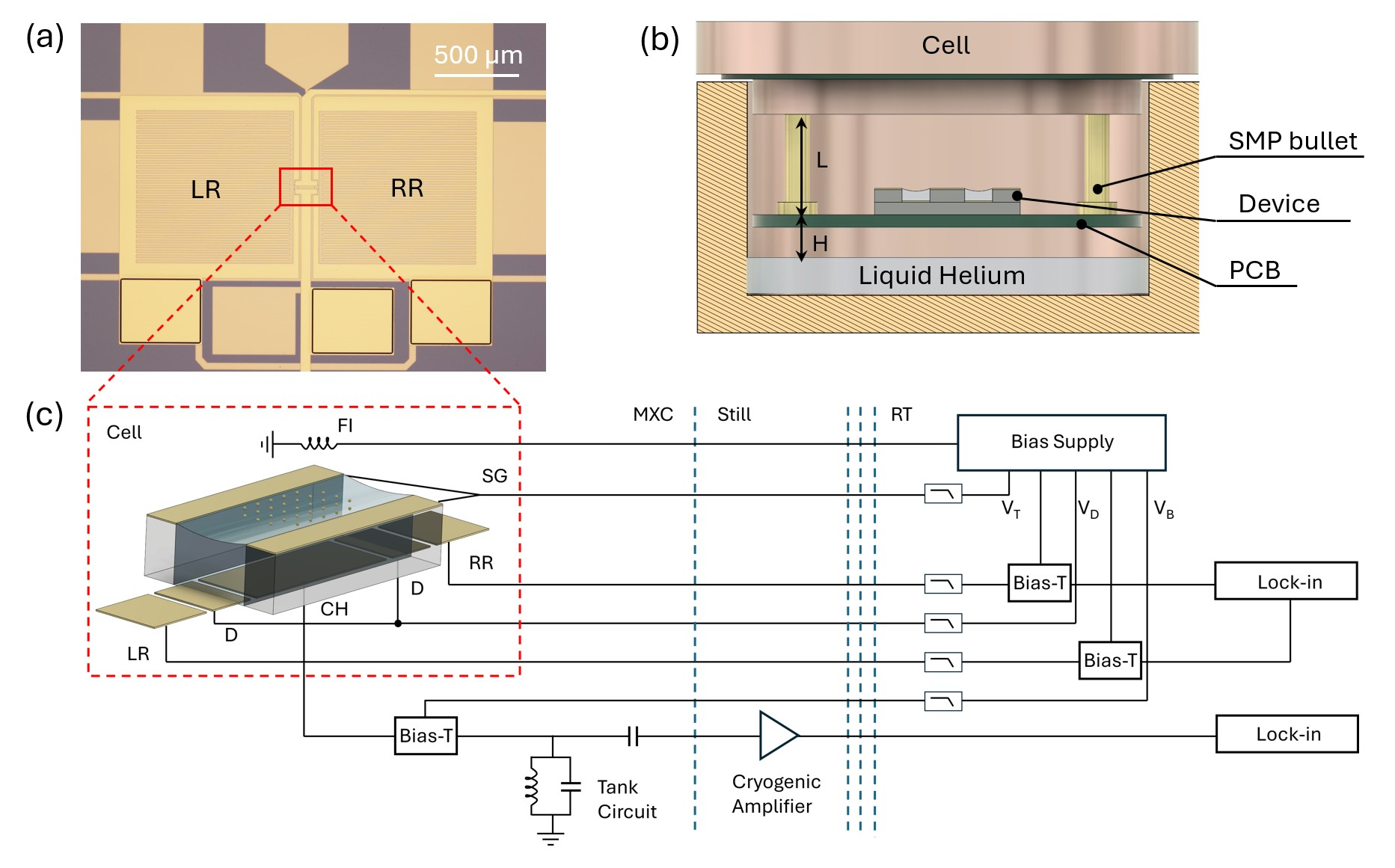}
\caption{(color online) (a) Microscopic image of the microchannel device used in the experiment. Left and right electron reservoirs (LR and RR, respectively) are connected by a central channel (inclosed in the red frame), whose details are shown in panel (c). (b) Schematic view of the experimental cell containing the device mounted on PCB. (c) Schematic view of the central channel and principle diagram of the measurement circuit.}
	\label{fig:1}	
	\end{figure}

 Our microchannel device has a standard geometry (see Fig.~\ref{fig:1}(a)) comprised of two channel arrays forming left and right reservoirs (LR and RR, respectively) for electron storage, which are connected by a single 140-$\mu$m long central channel having the width $w=10$~$\mu$m. This is a three-layer device consisting of two conducting 50~nm (gold) layers of separated by a layer of insulating photoresist whose thickness determines the channel height $h=1.5$~$\mu$m. The microchannel device is mounted on a printed-circuit board (PCB) inside a leak-tight experimental cell and is connected to the hermetic SMP connectors at the cell top by five SMP bullet adapters of length $L=20.3$~mm (see Fig.~\ref{fig:1}(b)). The level of bulk liquid helium is set below the PCB and  the channels of the device are filled with the superfluid helium by the capillary action, with the level of liquid in the channels determined by the balance of gravitational force and the surface tension. Assuming a round shape of the meniscus, the level of liquid at the center of the channel can be written as~\cite{koolNatComm2019} 

\begin{equation}
z_0=h - \frac{\rho g H}{2\sigma} \left( \frac{w}{2}\right)^2,
\label{eq:1}
\end{equation}

\noindent where $H$ is the distance between the level of bulk liquid and the device (see Fig.~\ref{fig:1}(b)), $\rho$ and $\sigma$ is the density and surface tension of liquid helium, respectively, and $g$ is the acceleration of a freely fallen body. From the amount of helium condensed inside the cell we estimate $H=6.3$~mm. The surface of liquid in the device is charged with electrons thermally emitted from a tungsten filament (FI) mounted above the device, while a positive bias voltage of 0.2~V is applied to the metal electrodes comprising the bottom of two reservoirs from a multichannel bias supply (Stahl-Electronics BS 1-12-10). After the charging, the surface of liquid in the central channel can be filled with eHe from the reservoirs and their density can be varied by adjusting the bias voltage applied to the three electrodes (D-CH-D) comprising the bottom of the channel (see Fig.~\ref{fig:1}(c)). Moreover, by applying a sufficiently large negative bias $V_{\rm D}$ to the side electrodes (D-D), we can isolate a fixed number of electrons in the channel above the middle electrode (CH). When this electrons are subject to a millimeter-wave (mm-wave) radiation introduced into the cell, they can be resonantly excited from the ground state to the first excited Rydberg states, with the corresponding transition frequency $f_{21}$ in the range 350-500 GHz, by Stark-tuning the energy levels with the dc bias voltage $V_{\rm B}$ applied to the bottom middle electrode (CH). Additionally, a negative bias voltage $V_{\rm T}$ can be applied to the top conducting layer (SG) of the channel to enhance confinement of eHe in the channel.

The Rydberg transition of eHe in the central channel is detected by the image-charge method using a resonant current amplifier developed earlier~\cite{belJLTP2024}. In this method, a change in the image charge induced by the excited electrons in a nearby electrode can be detected as an ac current excited in a connected electrical circuit~\cite{erikaPRL2019}. Our amplifier is comprised of a resonant tank (parallel LC) circuit one end of which is grounded, while another end is connected to the middle electrode at the bottom of the central channel (see Fig.~\ref{fig:1}(c)). eHe are excited by the pulse-modulated mm-wave radiation at the modulation frequency which coincides with the resonant frequency of the tank circuit ($\sim 1$ MHz), and an ac image-current induced by the excited electrons in the middle electrode is detected as a voltage drop across the large real impedance ($\sim 2.5$ M$\Omega$) of the tank circuit at resonance. This voltage signal is passed through a two-stage impedance-matching cryogenic amplifier and detected using a room-temperature lock-in amplifier referenced at the modulation frequency of the mm-wave radiation~\cite{belJLTP2024}. All data presented here were taken at the temperature cell $T\approx 250$~mK measured by a ruthenium oxide sensor attached to the cell.

\section{Results and discussion}
\label{sec:result}

\begin{figure}[htt]
	\includegraphics[width=1.0\textwidth]{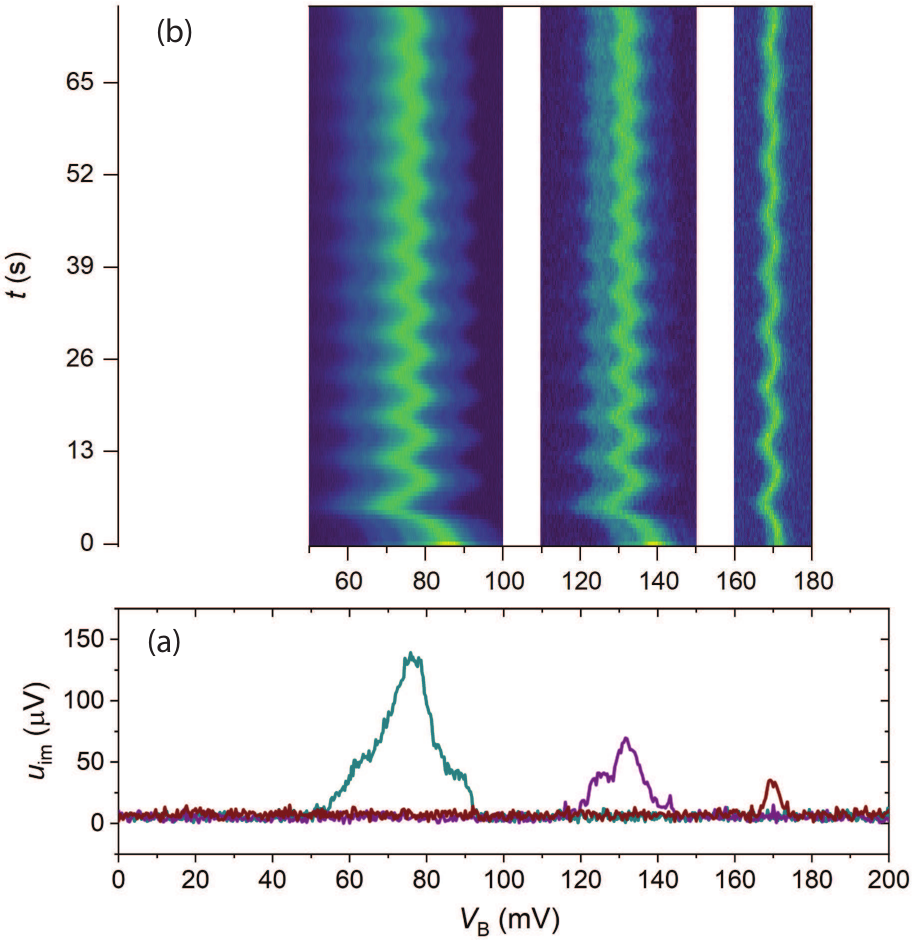}
\caption{(color online) (a) Stark spectra of the Rydberg transition for different fixed numbers of electrons trapped in the central microchannel of the device. The number of electrons decreases for three spectra from left to right. (b) Time evolution of the Stark spectra for the three cases considered in panel (a). The horizontal axis is the same as for the plot in panel (a).  All data are taken at $V_{\rm T}=-0.4$~V.}
	\label{fig:2}	 
\end{figure}

First, we consider a situation when a fixed number of eHe is confined in the central channel by applying a negative voltage $V_{\rm D}=-0.4$~V to the electrodes on the two sides of the channel (D-D). The Rydberg transition of these electrons is recorded at a fixed frequency of the pulse-modulated mm-wave radiation $f_0$ by varying the transition frequency of electrons $f_{21}$ with the bias voltage $V_{\rm B}$. Fig.~\ref{fig:2}(a) shows three exemplary Stark spectra recorded for different number of electrons confined above the middle electrode (CH). The measurements are done at the  mm-wave frequency $f_0=389.52$~GHz and the duty cycle of pulse modulation $D=15\%$ to avoid excessive heating of the experimental cell by the radiation. The voltage signal $u_{\rm im}$ is recorded by a lock-in amplifier with the integration time 0.8~s and the voltage $V_{\rm B}$ sweeping rate of 0.3~mV/s. It is observed that with the decreasing number of electrons the signal amplitude decreases and the transition spectrum shifts towards higher values of $V_{\rm B}$. Qualitatively, as the number of electrons decreases, the contribution to the perpendicular electric field $E_\perp$ exerted on eHe due to the positive image charges induced at the bottom electrode decreases, which must be compensated by increasing the positive bias voltage $V_{\rm B}$. A more qualitative analysis will be given later. 

The linewidth of the transition spectra shown in  Fig.~\ref{fig:2}(a) can be estimated by knowing the conversion coefficient between the transition frequency $f_{21}$ and the applied bias voltage $V_{\rm B}$. In the experiment, this coefficient is determined by changing the mm-wave frequency $f_0$ and recording the position of the Stark-shifted spectra. In the frequency range considered here, the dependance of $f_{21}$ on $V_{\rm B}$ was found to be almost linear with the slope $df_{21}/dV_{\rm B}\approx 0.4$~GHz/mV. The linewidth of the spectra observed in Fig.~\ref{fig:2}(a) on the order 10~GHz is expected to arise from the inhomogeneous broadening by a nonuniform electric field $E_\perp$ experienced by eHe in the channel~\cite{zouJLTP2022}. However, one should not exclude other sources of line broadening, in particular due to the time variation of the superfluid helium depth in the channel. Note that the typical time required to sweep the bias voltage $V_{\rm B}$ through the resonance is several minutes for the spectra shown in Fig.~\ref{fig:2}(a). Therefore, any variations in the transition spectrum on a faster timescale are expected to be averaged. Faster sweeping rate requires decreasing of the integration time of the lock-in amplifier, which degrades the signal-to-noise ratio. Instead, we fix the bias voltage $V_{\rm B}$ and record the lock-in signal as a function of time with the same integration time of 0.8~s. By repeating such measurements at different values of $V_{\rm B}$ we can observe the temporal dynamics of the Stark spectrum of eHe. Fig.~\ref{fig:2}(b) shows results of such a measurement for the same number of trapped eHe as in Fig.~\ref{fig:2}(a). This measurement reveals a time oscillation of the Stark spectra with the period of about 7~s. We found that such oscillations can be easily triggered by a slight heating of the experimental cell. In particular, we found that it is convenient to trigger oscillations by simply changing the duty cycle of the applied pulse-modulated radiation, which is mostly absorbed on the cell's inner wall during multiple reflection. For each time trace shown in Fig.~\ref{fig:2}(b), the duty cycle $D$ was switch to $D=0.4\%$ for the time duration $\Delta t=4$~s and then returned back to $0.15\%$ at $t=0$. In what follows, we will refer to this as the ``excitation'' pulse sequence. It was observed that the oscillation amplitude decreases with time after the excitation and the oscillations can persist for at least several minutes. 

\begin{figure}[htt]
\centering
	\includegraphics[width=1\textwidth]{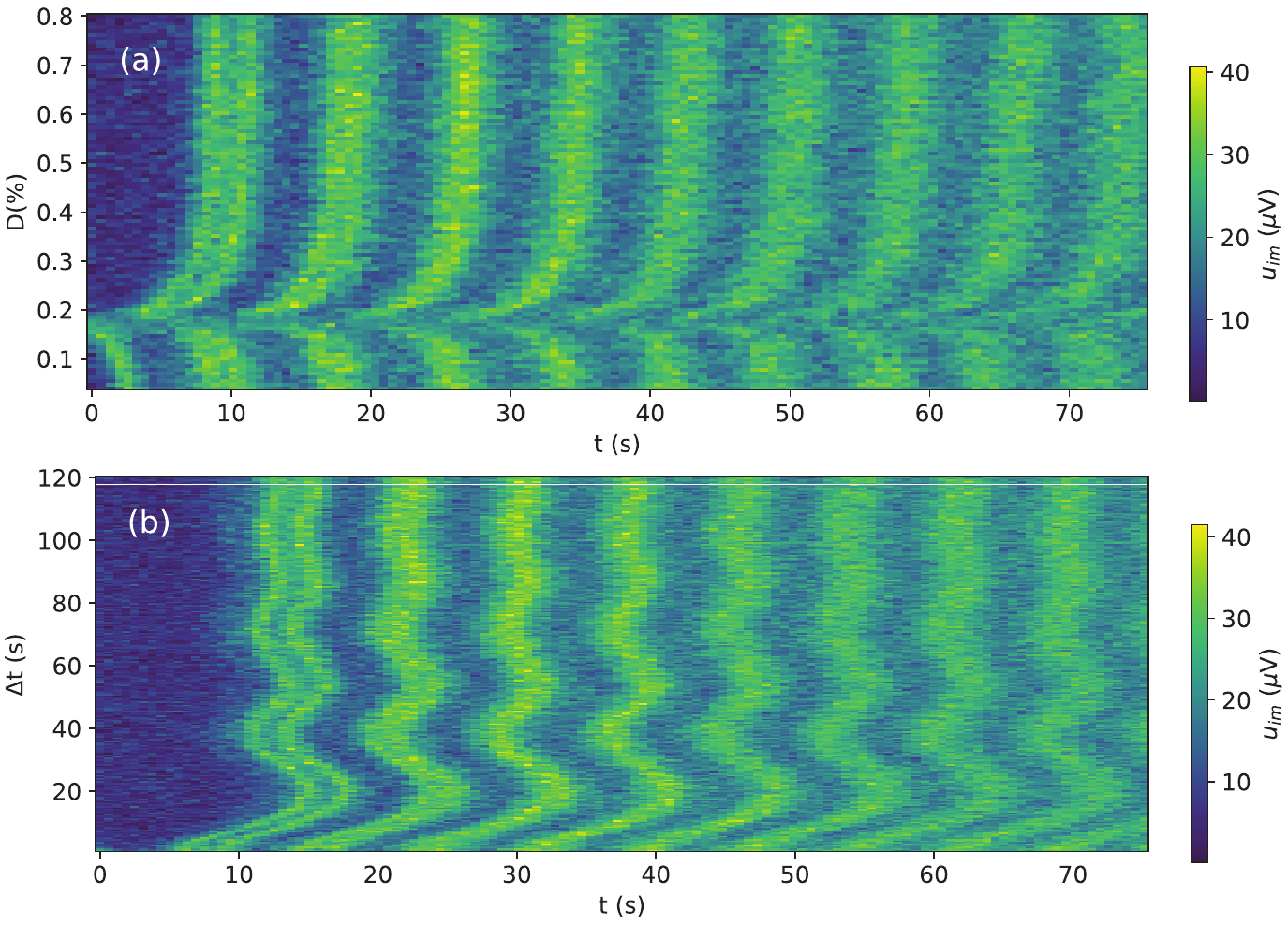}
\caption{(color online) (a) Time evolution of the image-charge signal of trapped eHe in response to the excitation pulse sequence at $t<0$ for different values of its duty circle $D$. (b) Time evolution of the image-charge signal for the same electrons as in (a) for different values of the excitation pulse duration $\Delta t$. All data are taken at $V_{\rm B}=168$~mV and $V_{\rm T}=-0.4$~V.}
	\label{fig:3}	 
\end{figure}

Fig.~\ref{fig:3}(a) shows time traces of the lock-in signal measured for the fixed number of electrons corresponding to the right-most resonance in Fig.~\ref{fig:2}(a) at a fixed value of $V_{\rm B}=168$~mV and different values of the duty cycle $D$ of the excitation pulse sequence. The value of $V_{\rm B}$ was chosen to be at the maximum slope of the $u_{im}$ versus $V_{\rm B}$ dependance for the Stark spectrum in Fig.~\ref{fig:2}(a). As for the data in Fig.~\ref{fig:2}(b), the excitation pulse sequence was applied for $\Delta t=4$~s, after which $D$ was changed back to to 0.15~\% at $t=0$. The time oscillations of the transition signal are clearly observed when the duty cycle $D$ of the excitation pulse sequence is not equal to $0.15\%$. The oscillations are not visible for $D=0.15\%$, when there is essentially no excitation sequence. This suggests that the oscillations are eventually damped under the steady-state pulse-modulated excitation. For $D<0.15\%$, the oscillations seem to appear a half-period earlier. For such values of $D$, the excitation pulse actually induces cooling of the cell rather than heating. This suggests that the phase of the oscillations depends on their excitation mechanism. Fig.~\ref{fig:3}(b) shows time traces measured with the same number of electrons, the fixed duty cycle of the excitation pulse sequences $D=0.26\%$ and for different time duration of the excitation sequence $\Delta t$. Note that the heating of the experimental cell induced by the excitation pulse increases with increasing $D$ and $\Delta t$ in Fig.~\ref{fig:3}(a) and Fig.~\ref{fig:3}(b), respectively. The data in both figures suggest that the onset time for the oscillations, which first increases with the heating, becomes relatively insensitive to its increase when the heating is sufficiently strong.  

\begin{figure}[htt]
\centering
	\includegraphics[width=0.8\textwidth]{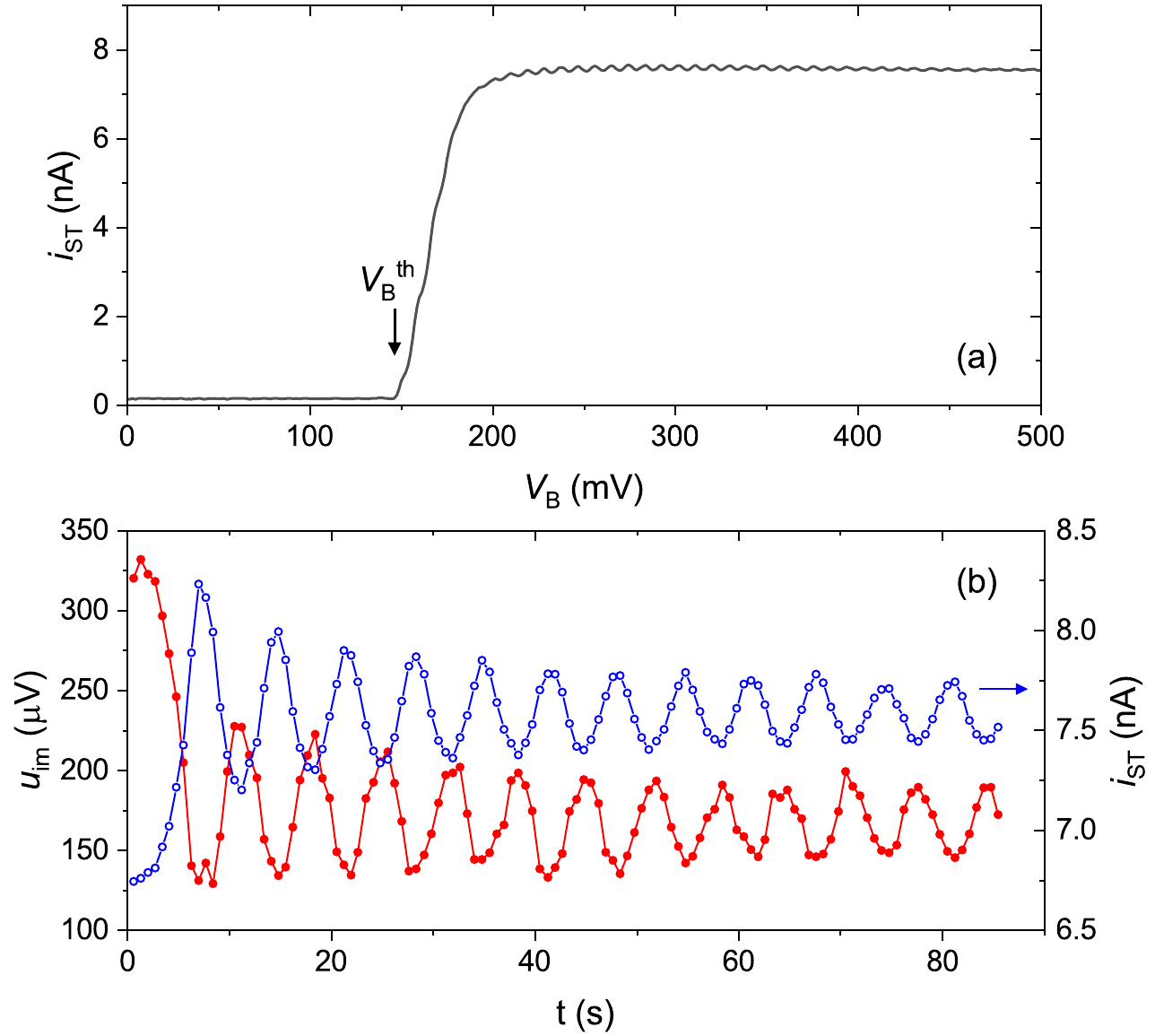}
\caption{(color online) (a) Sommer-Tanner signal proportional to the current of eHe through the central channel versus the bottom electrode voltage. The arrow indicates the threshold voltage $V_{\rm B}^{\rm th}$ below which the current is zero. (b) The image-charge (closed circles, left axis) and Sommer-Tanner (open cycles, left axis) signals of eHe measured simultaneously as a function of time after applying an excitation pulse sequence of the mm-wave radiation. eHe are tuned to the Rydberg resonance by setting $V_{\rm B}=294$~mV. All data are taken at $V_{\rm T}=-0.1$~V.}
	\label{fig:4}	 
\end{figure}

In an alternative configuration, we apply a positive voltage $V_{\rm D}=0.6$~V to the electrodes on the two sides of the central channel (D-D) thus allowing electrons from the reservoirs to enter the channel. In such a case, in addition to the Rydberg transition of eHe in the channel, we can also measure the AC current of electrons through the channel when electrons are driven between the two reservoirs by the standard Sommer-Tanner (ST) method~\cite{sommPRL1971}. For this purpose, we apply a 10~mV rms voltage excitation $v$ at the frequency $f_{\rm ST}=18$~kHz to the LR electrode and measure the current induced by the motion of eHe at the RR electrode using a lock-in amplifier (see Fig.~\ref{fig:1}(c)). Fig.~\ref{fig:4}(a) shows the dependance of a ST current $i_{\rm ST}$ on the bias voltage applied to the bottom middle electrode of the channel. Below the threshold value $V_{\rm B}^{\rm th}\approx 147$~mV, the current is zero (a small offset is due to the capacitive crosstalk between the reservoir electrodes) indicating that the central channel is empty of electrons. Above the threshold, the signal is rapidly rising, which indicates that eHe fill the channel. Surprisingly, by using the excitation pulse sequence similar to the previously described, we are able to observe a similar temporal dynamics of the measured ST signal of eHe. Fig.~\ref{fig:4}(b) shows an example of measured $i_{\rm ST}$ (open cycles) obtained by applying the same pulse sequence of the mm-wave radiation as for the data presented in Fig.~\ref{fig:2}(b). The image-charge signal of eHe in the channel $u_{\rm im}$ measured simultaneously with the ST signal is also shown (closed cycles). In this measurement, the Rydberg transition of eHe in the channel is set to be at resonance with the radiation by setting $V_{\rm B}=294$~mV. There is a clear correlation between the time oscillations of these two quantities, which unambiguously demonstrates that they share a common origin.

Considering the time scale and quality factor of these oscillations, we believe it is reasonable to suggest that they originate from the oscillations of the superfluid helium level in the channel. To estimate the variation of the liquid level from the experimental data shown in Fig.~\ref{fig:1}, we use a simplified model according to which a uniformly charge surface of liquid helium in the channel forms a parallel-plate capacitor with the bottom electrode located at the distance $z$ below the surface, where $z$ is the mean depth of liquid helium in the channel. In accordance with this model, the perpendicular electric field experienced by eHe in the channel can be written as~\cite{zouJLTP2022} 

\begin{equation}
E_\perp = \frac{V_{\rm B}-V_0}{z} - \frac{en_s}{2\varepsilon_0},
\label{eq:2}
\end{equation}

\noindent where $V_0$ is the electric potential of the charged surface of liquid in the channel, $n_s$ is the areal density of eHe in the channel, and $\varepsilon_0$ is the electrical permittivity of vacuum. Note that the above expression includes the contribution to $E_\perp$ from both the applied bias voltages and the positive image charge induced in the conducting electrodes by eHe. Using the same model, we can write the relation between $V_0$ and $n_s$ as~\cite{zouJLTP2022}

\begin{equation}
n_s=\frac{\varepsilon\varepsilon_0}{\alpha e z}\left( \alpha V_{\rm B} + \beta V_{\rm T} - V_0\right),
\label{eq:3}
\end{equation}

\noindent where $V_{\rm T}$ is the bias voltage applied to the top conducting layer of the channel, $\alpha$ and $\beta=1-\alpha$ are relative contributions to the total capacitance of the charged surface due to the bottom and top conducting layer, respectively, and $\varepsilon$ is the dielectric constant of liquid helium. From the above equations we can find a relation between the variation of the liquid depth in the channel $\delta z$ and the variation of the bias voltage $\delta V_{\rm B}$ while eHe are maintained at the transition resonance (that is fixed $E_\perp$). For example, in the case of a fixed electron density as in Fig.~\ref{fig:2}(b), we easily obtain $\delta V_{\rm B} = \delta z(V_{\rm B} - V_{\rm T})/z$. With $\delta V_{\rm B} \approx 10$~mV and $z\approx h$, we obtain an estimate $\delta z \approx 250$, 226 and 180~nm for the three cases shown in Fig.~\ref{fig:2}(b). Note that for sufficiently low electron density, as for the right-most plot in Fig.~\ref{fig:2}(b), the above parallel-plate capacitor model is not adequate, thus giving an underestimation for $\delta z$. In general, we find that the variation of helium depth in the channel is about $20\%$ at the onset of oscillations.  

The above explanation also agrees with the observed oscillations of the ST signal shown in Fig.~\ref{fig:4}(b). According to the data shown in~Fig.~\ref{fig:2}(b), the Rydberg resonance of eHe immediately after the excitation of oscillations ($t=0$) start shifting towards the lower values of $V_{\rm B}$. This suggests that the depth of liquid $z$ starts decreasing, thus enhancing the electric field component due to the image charge. Simultaneously, we observe an increase in the ST signal of eHe current through the central channel, see Fig.~\ref{fig:4}(b). In the capacitive-coupling regime considered in Fig.~\ref{fig:4}, the electron current is given by $i_{\rm ST}\approx 2\pi f_{\rm ST}Cv$, where $C$ is the capacitance between the charged surface of liquid in the device and the bottom reservoir electrodes. The latter is inversely proportional to the helium depth $z$, which is in a complete agreement with the observed increase of $v_{\rm ST}$ immediately after the excitation and a $\pi$ difference in phase between the oscillations of $u_{\rm im}$ and $v_{\rm ST}$. Also, the amplitude of $v_{\rm ST}$ oscillations in Fig.~\ref{fig:4}(b) is in an order-of-magnitude agreement with the above estimation of $\delta z$. We also note that a weakly damped oscillation in the ST signal could be clearly observed after the charging of the helium surface by the thermal emission from the filament. The charging is typically accompanied by a significant heating of the cell due to the Joule heating of the filament. It was found that such oscillations of the ST signal has a similar frequency as discussed above and could last for thousands of seconds.                

Finally, we discuss a possible origin of these oscillations. Helium surface fluctuations due to the mechanical vibration of the cryostat has been considered in details in Refs.~\cite{koolNatComm2019,beysJLTP2022}. The typical amplitude of helium depth oscillations observed in a channel with similar dimensions as considered here was estimated to be about 1~nm, which is two orders of magnitude smaller than considered here. It is unlikely that the mechanical vibrations of our system can lead to the observed effect. The flow of superfluid helium in the cell filling line due to temperature gradients can give rise to some unwanted effects, such as oscillations in the temperature of the different stages of a cryostat~\cite{casRSI2024}. We did not observe any oscillations in the temperature of the experimental cell in the considered frequency range. The redistribution of superfluid helium in the filling line can also potentially lead to the variation of the bulk helium level $H$ in the cell. However, according to Eq.~(\ref{eq:1}) this would produce a negligible effect on the helium level in the channel. We are inclined to believe that the oscillation of helium level observed in the experiment is due to the motion of supefluid helium film inside the experimental cell. It is well known that when a container of liquid helium is connected by a fine slit or a capillary with a helium bath, the liquid meniscus in the container can undergo slow and weakly damped oscillations due to the inertia of the superfluid flow in the capillary~\cite{allProc1936,robPR1951}. A similar phenomenon can occur through a superfluid helium film~\cite{atkProc1950}. In our setup, the liquid helium in the channels is connected with the bulk helium by a superfluid film covering the SMP bullet adapters and the walls of the cell (see Fig.~\ref{fig:1}(b)). At equilibrium, the helium level in the channel is set by the balance between the gravitational force and the surface tension according to Eq.~(\ref{eq:1}). If the level is lowered, the excess superfluid helium must flow back to the bulk liquid below the chip via the film covering the surface of the bullet adapters. The inertia of this flow can lead to a damped oscillation of the helium level observed in the experiment. An order-of-magnitude estimation for the oscillation frequency can be made by considering the cross-section of the suprfluid film $S$ and the total area of liquid helium in the microchannels $A$ according to $\omega=2\pi f_{\rm osc} = \sqrt{gS/(LA)}$~\cite{allProc1936,robPR1951,atkProc1950}. Assuming an average film thickness of 30~nm covering all five bullet adapters, each having a diameter of 3.6~mm, and the estimated total area of microchannels $A= 1.2 $~mm$^2$, we obtain $f_{\rm osc}\approx 0.4$~Hz. Considering extreme simplicity of the above estimation, the agreement with the experimentally observed frequency of 0.14~Hz is satisfactory.           

\section{Conclusion}
\label{sec:conclusion}

In summary, we observed sub-hertz weakly damped oscillations of the liquid helium depth in a microchannel device for electron trapping. The oscillations are revealed by observing the temporal dynamics of the electron transition between the Rydberg states of their quantized out-of-plane motion. Such observation is possible because the variation of helium depth affects the Rydberg transition spectrum via the Stark shift due to the applied bias voltages and the image charges induced in the conducting electrodes of the device. We attribute the observed oscillations of the helium depth in the channels to the inertia of the helium mass transfer through the superfluid film inside the cell. We note that the above hypothesis can be further tested by varying various experimental parameters, such as the bulk helium depth $H$, the microchanel area $A$, etc. Also, it is desirable to establish a proper theoretical framework capable to describe the proposed mechanism of oscillation in the geometry of our experimental setup. Clarifying the origin of such oscillations would be important for seeking proper countermeasures to reduce or completely eliminate such an effect.              

\begin{acknowledgements}
The work was supported by an internal grant from the Okinawa Institute of Science and Technology (OIST) Graduate University and the Grant-in-Aid for Scientific Research (Grant No. 23H01795 and 23K26488) KAKENHI MEXT. 
\end{acknowledgements}




\end{document}